\documentclass[english,aps,a4paper,preprint,groupedaddress,floatfix]{revtex4}%
\usepackage[T1]{fontenc}
\usepackage[latin1]{inputenc}
\usepackage{color}
\usepackage{graphicx}
\usepackage{epsfig}
\usepackage{babel}
\usepackage{amsmath}
\usepackage{amsfonts}
\usepackage{amssymb}%
\newcommand{\be}{\begin{eqnarray}}
\newcommand{\ee}{\end{eqnarray}}
\newcommand{\ba}{\begin{array}}
\newcommand{\ea}{\end{array}}

\begin{document}
\title{On strange SU(3) partners of $\Theta^{+}$}
\author{Klaus Goeke$^{1}$}
\email{Klaus.Goeke@tp2.ruhr-uni-bochum.de}
\author{Maxim V. Polyakov$^{1,2}$}
\email{Maxim.Polyakov@tp2.ruhr-uni-bochum.de}
\author{Michal Praszalowicz$^{3}$}
\email{michal@if.uj.edu.pl} \affiliation{$^{1}$~Institut f\"ur
Theoretische Physik II, Ruhr-Universit\"at Bochum, D-44780,
Germany, $^{2}$~Petersburg Nuclear Physics Institute,
188 300 Gatchina, St. Petersburg, Russia,\\  }
\affiliation{$^{3}$~M.Smoluchowski Institute of Physics,
Jagellonian University, ul.Reymonta 4, 30-059 Krak{\'o}w, Poland.}
\date{\today }

\begin{abstract}
We propose a scenario in which Roper octet can mix with a putative
antidecuplet of exotic baryons and predict the properties of its strange
members. We show that $1795\;\text{MeV} < M_{\Sigma_{\overline{10}}%
}<1830\;\text{MeV}$ and $1900\;\text{MeV} < M_{\Xi_{\overline{10}}%
}<1970\;\text{MeV}$. We also estimate total widths: $10\; \text{MeV}%
<\Gamma_{\Sigma_{\overline{10}}}<30\; \text{MeV}$ and $\Gamma_{\Xi
_{\overline{10}}}\sim10 \; \text{MeV}$ and branching ratios for different
decay modes.

\end{abstract}
\maketitle

\section{Introduction}

\label{sect:introduction}

At the end of 2002 Japanese collaboration LEPS at Spring-8 \cite{Nakano} and
the bubble chamber experiment DIANA at ITEP, Moscow \cite{Dolgolenko},
announced evidence of a strange baryon, called $\Theta^{+}$, whose quantum
numbers cannot be constructed from 3 quarks. This truly exotic state:
($uudd\bar{s}$) has been anticipated already by the founders of the quark
model, however it was believed to be rather heavy (1800 -- 1900 MeV) and wide.
The excitement created by the findings of LEPS and DIANA was due to the low
mass of the putative $\Theta^{+}$, of the order of 1530 MeV, and a very small
width. Such properties, however, are natural in chiral models where the
antistrange quark is excited in form of a chiral field rather than as a
constituent quark. Early predictions of the $\Theta^{+}$ mass in different
versions of the chiral models were very close to the experimental numbers of
LEPS and DIANA \cite{BieDo,Praszalowicz:2003ik}, and moreover the width, as
estimated by Diakonov, Petrov and Polyakov in 1997, was very small \cite{dia}.

The discovery by LEPS and DIANA triggered both experimental and theoretical
activity in a somewhat extinct field of hadron spectroscopy. Present
experimental situation is, however, rather confusing. Many collaborations
announced the existence of $\Theta^{+}$ in different experimental setups, but
the comparable number of searches ended with null result. Furthermore,
dedicated high statistics runs with CLAS detector at CEBAF did not show any
signal of $\Theta^{+}$ contradicting previous findings of the same
collaboration \cite{CLAS-no}. On the other hand LEPS has confirmed its
original result \cite{Hotta:2005rh,Nakano_confirm}. Also high statistics
analysis by DIANA turned out to be positive \cite{Barmin_confirm}. We refer
the reader to recent reviews of the experimental situation
\cite{Nakano:2005ut,Burkert:2005ft,Danilov:2008zza}. For the purpose of the
present study we assume that $\Theta^{+}$ exists with a mass equal 1540 MeV
and total width $\Gamma<1$ MeV.

\begin{figure}[t]
\begin{centering}
\includegraphics[scale=2.0]{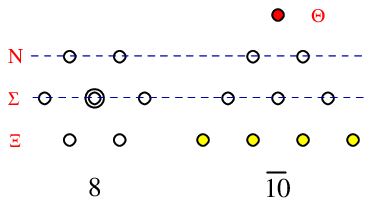}
\par\end{centering}
\caption{SU$_{\mathrm{fl}}(3)$ weight diagrams for octet and
antidecuplet.
States that can mix lie on dashed lines.}%
\label{fig:multiplets}%
\end{figure}

One of the immediate consequences of a possible existence of $\Theta^{+}$ is
the existence of the whole SU(3) multiplet: $\overline{10}$ (see
Fig.\ref{fig:multiplets}). Indeed, NA49 experiment at CERN \cite{Alt:2003vb}
announced discovery of another exotic state, namely $\Xi_{\overline{10}}%
^{--}(1860)$. Unfortunately the searches of $\Xi_{\overline{10}}$ by other
groups have not confirmed the results of NA49.

Apart from truly exotic states that cannot be constructed from three quarks,
antidecuplet contains cryptoexotic states that are primarily built from 5
quarks, however their quantum numbers can be constructed from three quarks as
well. These are nucleon-like states ($N_{\overline{10}}$) and $\Sigma-$like
states ($\Sigma_{\overline{10}}$) that are the subject of the present paper.
The interpretation of these states is not well understood: one may try to
associate them with some known resonances, as it was done in the case of
$N_{\overline{10}}$ in the original paper of Diakonov, Petrov and Polyakov for
example \cite{dia}, or one may postulate the existence of new, yet
undiscovered resonances with nucleon or $\Sigma$ quantum numbers. In this
paper we follow the latter approach trying to predict the range of masses and
widths for cryptoexotic $\Sigma_{\overline{10}}$ and also $\Xi_{\overline{10}%
}$ states.

\renewcommand{\baselinestretch}{0.5}
\renewcommand{\baselinestretch}{1.0}

In the Chiral Quark-Soliton model ($\chi$QSM) the spin-parity quantum numbers
of the antidecuplet members are unambiguously predicted to be $J^{P}=\frac
{1}{2}^{+}$ \cite{dia}, so that $N_{\overline{10}}$ and $\Sigma_{\overline
{10}}$ are predicted to be $P_{11}$ resonances. One of the striking properties
of $N_{\overline{10}}$ is that it can be excited by an electromagnetic probe
from the neutron target much stronger than from the proton one~\cite{max}. The
photoexcitation of charged isocomponent of $N_{\overline{10}}$ is possible
only due to SU$_{\mathrm{fl}}(3)$ violation; therefore its suppression by a
factor $\sim1/10$ in the amplitude is expected.

The existence of a new nucleon resonance with the
mass near $\sim 1700$ MeV was suggested in Refs.~\cite{dia1,str}.
The authors of Ref.~\cite{dia1} used the Gell-Mann--Okubo mass
relations in the presence of mixing, in order to predict the mass
of this new nucleon resonance. As an input for the
Gell-Mann--Okubo  mass formula the authors of Ref.~\cite{dia1}
used the mass of the $\Xi_{\overline{10}}^{--}$ baryon reported by the NA49
collaboration \cite{Alt:2003vb}. In Ref.~\cite{str}, in order to constrain the
mass of this possible new narrow $N^*$, the modified PWA of $\pi
N$ scattering data was employed. It was found that the easiest way
to accommodate a narrow $N^*$ is to set its mass around $1680$~MeV
and quantum numbers to $P_{11}$ ($J^P=\frac 12^+$). In the same
paper the width of the possible $N^*$ was analyzed in the
framework of $\chi$QSM. It was found that the width of new N$^*$
is in the range of tens of MeV with very small $\pi N$ partial width of
$\Gamma_{\pi N}\leq 0.5$~MeV \cite{str}. One should stress that the
decay to $\pi N$ is not suppressed in the SU$_{\mathrm{fl}}(3)$ limit and it
can be made small only if the symmetry violation is taken into account. It
follows that the preferred decay channels are $\eta N$, $\pi\Delta$ and
$K\Lambda$ ~\cite{dia,str,michal1,michal2,guz2,Ledwig:2008rw}.

The search of the new nucleon
has been performed in the $\eta$ photoproduction on
the neutron at GRAAL~\cite{Kuznetsov04,gra1}.
The narrow peak in the quasi-free neutron cross section
 and in the $\eta n$ invariant mass spectrum has been observed.
The original observation of Refs.~\cite{Kuznetsov04,gra1} has been
recently confirmed by two other groups: CBELSA/TAPS~\cite{kru} and
LNS-Sendai~\cite{kas}. All three experiments found an enhancement
in the quasi-free cross-section on the neutron. In addition, the GRAAL and CBELSA/TAPS groups
have observed a narrow peak in the $\eta n$ invariant mass
spectrum at $1680-1685$~MeV. The corresponding data can be explained
by existence of a new narrow resonance \cite{Azimov:2005jj,tia1,kim}.
 One should, however, stress
that the part of above mentioned experimental results may have a different
interpretation that does not require to postulate a new narrow nucleon
resonance \cite{Shklyar:2006xw,Anisovich:2008wd,Doring:2009qr}.

Further evidence for a new narrow nucleon resonance came from the
analysis of the $\Sigma$ beam asymmetry in $\eta$ photoproduction on the proton
\cite{Kuznetsov:2008hj,Kuznetsov:2008gm,Kuznetsov:2007dy,Kuznetsov:2008ii}.
In these papers the narrow structure in the $\Sigma$ beam asymmetry around
invariant mass $\sim 1685$~MeV has been observed. That structure can be
interpreted as the contribution of a narrow nucleon resonance with the
mass 1685~MeV, total width $\leq 25$~MeV and the photocoupling to the proton
much smaller than to the neutron; the properties that are expected for the
nonstrange partner of $\Theta^+$\cite{max,dia,str,michal1,michal2,guz2,Ledwig:2008rw}.
The properties of possible new narrow nucleon resonance that crystalized out
recent experiments on $\eta$ photoproduction are summarized in Ref.~\cite{Kuznetsov:2008ii}.
Throughout this paper we shall assume that the new $N(1685)$ nucleon resonance
exists with  total width below 25 MeV.

One of the striking and to some extent counterintuitive properties of
$\Theta^{+}$ is its small width. In particular, DIANA \cite{Barmin_confirm}
that has doubled the statistics of their formation experiment $K^{+}%
n(\mathrm{Xe})\rightarrow K^{0}p$ as compared to the original result
reported in Ref.\cite{Dolgolenko}, claims $M_{\Theta}=1537\pm2\,\mathrm{MeV}$
and the width $\Gamma_{\Theta}=0.39\pm0.10\,\mathrm{MeV}$ (with possible
systematic uncertainties). The only other available formation experiment with
the secondary kaon beam at BELLE sets an upper limit $\Gamma_{\Theta
}<0.64\,\mathrm{MeV}$ (at a 90\% confidence level)~\cite{Abe:2005gy} which is
beyond the above value. Also the reanalysis of the old $KN$ scattering
data~\cite{Arndt:2003xz} shows that there is room for the exotic resonance
with a width below 1~MeV.

The small width implies that the coupling $g_{\Theta NK}$ is at least an order
of magnitude smaller than $g_{\pi NN}\!\approx\!13$. The small value of
$g_{\Theta NK}$ appears naturally in a relativistic field-theoretic approach
to baryons, allowing for a consistent account for multi-quark components in
baryons; in particular in Refs.~\cite{dia5,Lorce} an upper bound $\Gamma
_{\Theta}\approx2\,\mathrm{MeV}$ has been obtained without any parameter
fixing. Recent calculation of the $\Theta^+$ width in $\chi$QSM \cite{Ledwig:2008rw}
also gave small width of $0.71$~MeV.
The width below 1~MeV also comes out from the parameter-free QCD sum rules
analysis~\cite{Oganesian:2006ug}.

In any case the small value of $g_{\Theta NK}$ is related to the small value
of $G_{\overline{10}}$ reduced matrix element that is responsible for the
direct decay $\overline{10}\rightarrow8$. Indeed%
\begin{equation}
g_{\Theta NK}=G_{\overline{10}}+\sin\alpha\,H_{\overline{10}} \label{gThN}%
\end{equation}
where $\alpha$ is the mixing angle between the nucleon-like states in octet
and antidecuplet, and $H_{\overline{10}}$ denotes the transition reduced
matrix element $\overline{10}\rightarrow\overline{10}$ (see Tab. \ref{tablematel}
for definitions). For ordinary baryons we expect terms like $\sin
\alpha\,H_{\overline{10}}$ ($\sin\alpha$ being of the order of $m_{s}$) to be
small in comparison with the leading term. For antidecuplet both terms
$G_{\overline{10}}$ and $\sin\alpha\,H_{\overline{10}}$ are small and
comparable in magnitude. They may add or cancel depending on the decay channel
(for $\Theta^{+}$ we have only two equal decay channels $\Theta^{+}\rightarrow
NK$) violating completely the SU$_{\text{fl}}(3)$ relations between the decay
couplings. In this way one is able to explain the suppression of $\pi N$ decay
channel in $N_{\overline{10}}$ decays mentioned above, and also the existence
of $\overline{10}\rightarrow10$ transition that is forbidden in the
SU$_{\text{fl}}(3)$ limit. We see therefore that in the case of antidecuplet
mixing is an important ingredient primarily to understand the decays, but also
the masses \cite{dia1,str,michal2,guz2,Pakvasa:2004pg}.

\begin{figure}[t]
\begin{centering}
\includegraphics[scale=2.3]{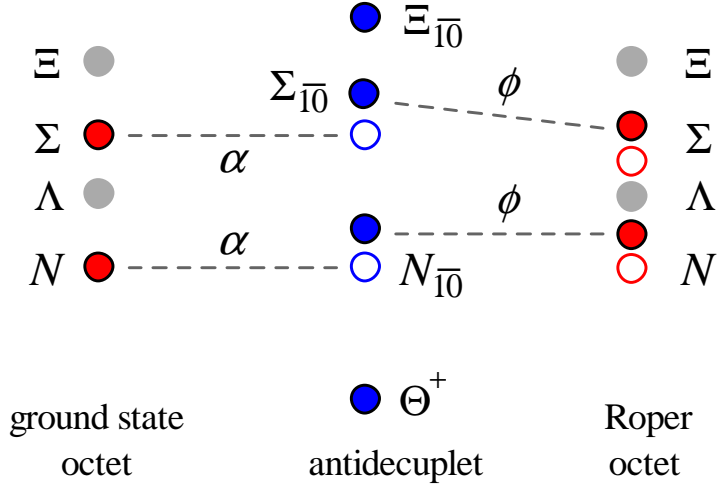}
\par\end{centering}
\caption{Definition of mixing angles for nucleon-like states and $\Sigma$-like
states. Full circles denote physical states, open circles in the case of Roper
octet and $\overline{10}$ correspond to the GMO states that undergo further
mixing with angle $\phi$. Grey circles correspond to particles not considered
in the present paper.}%
\label{fig:angles}%
\end{figure}

Unfortunately, at least at the first sight, there is large arbitrariness as
far as mixing angles and transition matrix elements are concerned. Here we
shall try to constrain them from the existing data. If the data are not
available we shall make estimates based on $\chi$QSM. We shall consider mixing
of the ground state octet ($8_{1}$), the octet of the $N(1440)$ Roper
resonance ($8_{2}$) and antidecuplet. In the SU$_{\text{fl}}(3)$ limit all
states in the ground state octet and in antidecuplet are separately degenerate
in mass. When $m_{s}$ corrections are switched on the masses split and take
values given by Gell-Mann--Okubo (GMO)\ mass formulae. At the same time the
wave functions become mixtures of the original representation (8 or
$\overline{10}$) and all other allowed SU$_{\text{fl}}(3)$ representations
that appear in the tensor product $8\otimes8$ or $8\otimes\overline{10}$. This
introduces $8-\overline{10}$ mixing for nucleon-like and $\Sigma$-like states
characterized by angle $\alpha$ (note that due to the accidental equality of
the pertinent SU(3) Clebsch-Gordan coefficients mixing angles between $N$- and
$\Sigma$-like states are equal in the leading order in $m_s$).
We assume that ground state octet GMO states
correspond to the physical states, which is true with 0.5\% accuracy
\cite{dia1}. We assume next that antidecuplet GMO states undergo further
mixing with the Roper octet. For Roper octet GMO mass formulae work with much
worse accuracy of approximately 3\% \cite{dia1}, so there is a need for
additional mixing. Again both for $N$- and $\Sigma$-like states the mixing
angle $\phi$ is the same. This is depicted in Fig.\ref{fig:angles}. Throughout
this paper we take into account only these two mixings which we generally
believe to be small.

\renewcommand{\baselinestretch}{0.5}
\renewcommand{\baselinestretch}{1.0}

Our analysis is based on the following assumptions concerning antidecuplet. We
assume $M_{\Theta^{+}}=1540$ MeV and $\Gamma_{\Theta^{+}}<1$ MeV. We follow
analysis of Refs.~\cite{Kuznetsov:2008hj,Kuznetsov:2008ii} assuming that $N_{\overline{10}}$ is a
new resonance with mass $1685$ MeV and total width $\Gamma_{N\overline{_{10}}%
}<25$ MeV. We also assume hierarchy of the branching ratios that is described
in more detail in Sect.~\ref{sect:where}. With these assumptions we are able to provide
limits on the mixing angles both for the nucleon-like and $\Sigma$-like
states. We find a small region in the space of mixing angles where
the required properties
of $N_{\overline{10}}$ are reproduced. For these allowed angles we calculate
masses of $\Sigma_{\overline{10}}$ and $\Xi_{\overline{10}}$ and their decay patterns.

The paper is organized as follows. In the next Section we discuss general
formulae for quantum mixing and define mixing angles and wave functions used
throughout this paper. In Sect.\ref{sect:decays} we calculate the decay
constants in the presence of mixing. We define reduced matrix elements and
discuss their hermiticity properties. Section \ref{sect:where} contains
numerical results of our analysis. Finally in Sect.\ref{sect:conclusions} we
briefly summarize our results and present conclusions.

\section{Masses in the presence of mixing}

\label{sect:masses}

\subsection{Two state mixing}

\label{sect:two}

Before we discuss three state mixing, let us recall the formulae for two state
mixing, that can be found for example in Ref.\cite{dia1}. Consider
perturbation hamiltonian ($M_{2}>M_{1},$ $V>0$) where $V\sim m_{s}$:%
\begin{equation}
H^{\prime}=\left[
\begin{array}
[c]{cc}%
M_{1} & -V\\
-V & M_{2}%
\end{array}
\right]  . \label{Hprim}%
\end{equation}
Here we have chosen "$-$" sign in front of $V$ in order to be in agreement
with the sign dictated by the $\chi$QSM. Let us consider for the moment
$\overline{10}-8$ mixing. In that case hamiltonian (\ref{Hprim}) represents
mixing between nucleon states or $\Sigma$ states (with different entries for
each case). Exotic states $\Theta^{+}$ and $\Xi_{\overline{10}}$
remain unmixed. It is
important to note here that the SU(3) Clebsch-Gordan coefficients are
identical for $N$ and $\Sigma$ states and therefore $V$ is the same.

Introducing%
\begin{equation}
\delta M=M_{2}-M_{1},\quad\Delta=\sqrt{\delta M^{2}+4V^{2}}%
\end{equation}
we get the following mass eigenvalues%
\begin{equation}
M_{1,2}^{\text{phys}}=\frac{1}{2}\left(  M_{1}+M_{2}\pm\Delta\right)
\end{equation}
and eigenvectors%
\begin{align}
\left\vert 1^{\text{phys}}\right\rangle  &  =\left\vert 1\right\rangle
\frac{2V}{\sqrt{2\Delta(\Delta-\delta M)}}+\left\vert 2\right\rangle
\sqrt{\frac{\Delta-\delta M}{2\Delta}},\nonumber\\
\left\vert 2^{\text{phys}}\right\rangle  &  =\left\vert 1\right\rangle
\frac{-2V}{\sqrt{2\Delta(\Delta+\delta M)}}+\left\vert 2\right\rangle
\sqrt{\frac{\Delta+\delta M}{2\Delta}}. \label{eigv2}%
\end{align}
The mixing angle $\alpha$
\begin{equation}
\left[
\begin{array}
[c]{c}%
\left\vert {}\right.  1^{\text{phys}}\left.  {}\right\rangle \\
\left\vert {}\right.  2^{\text{phys}}\left.  {}\right\rangle
\end{array}
\right]  =\left[
\begin{array}
[c]{rc}%
\cos\alpha & \sin\alpha\\
-\sin\alpha & \cos\alpha
\end{array}
\right]  \left[
\begin{array}
[c]{c}%
\left\vert 1\right\rangle \\
\left\vert 2\right\rangle
\end{array}
\right]  \label{rotation}%
\end{equation}
is conveniently defined through the second equation (\ref{eigv2})%
\begin{equation}
\tan\alpha=\frac{2V}{\Delta+\delta M}\simeq\frac{V}{\delta M}\simeq\frac
{V}{\overline{M}_{2}-\overline{M}_{1}}. \label{angle_pt}%
\end{equation}
The last approximation consists in approximating $\delta M$ by the difference
of the mean multiplet masses in the spirit of the first order perturbation
theory in $m_s$. Then Eq.(\ref{eigv2}) implies%
\begin{align}
M_{1}^{\text{phys}}  &  =\frac{1}{\cos2\alpha}\left(  M_{1}\cos^{2}%
\alpha-M_{2}\sin^{2}\alpha\right)  ,\nonumber\\
M_{2}^{\text{phys}}  &  =\frac{1}{\cos2\alpha}\left(  M_{2}\cos^{2}%
\alpha-M_{1}\sin^{2}\alpha\right)  . \label{physmass}%
\end{align}

\subsection{Three state mixing}

\label{sect:three}

Let us consider three states belonging to ordinary octet -- $8_{1}$, "Roper"
octet -- $8_{2}$ and antidecuplet $\overline{10}$, with unperturbed masses
satisfying%
\begin{equation}
M_{8_{1}}<M_{8_{2}}<M_{\overline{10}}. \label{unpertmas}%
\end{equation}
The vector of nucleon-like (or $\Sigma$-like) states%
\begin{equation}
\left[
\begin{array}
[c]{l}%
\left\vert 8_{1}\right\rangle \\
\left\vert 8_{2}\right\rangle \\
\left\vert \overline{10}\right\rangle
\end{array}
\right]  \label{vector}%
\end{equation}
is a subject of mixing by the orthogonal matrix%
\begin{equation}
\mathcal{O}=\left[
\begin{array}
[c]{ccc}%
\cos\theta & \sin\theta & 0\\
-\sin\theta & \cos\theta & 0\\
0 & 0 & 1
\end{array}
\right]  \left[
\begin{array}
[c]{ccc}%
1 & 0 & 0\\
0 & \cos\phi & \sin\phi\\
0 & -\sin\phi & \cos\phi
\end{array}
\right]  \left[
\begin{array}
[c]{ccc}%
\cos\alpha & 0 & \sin\alpha\\
0 & 1 & 0\\
-\sin\alpha & 0 & \cos\alpha
\end{array}
\right]  . \label{Omatrix}%
\end{equation}
Throughout this paper we neglect Roper-ground octet mixing and set $\theta=0
$. Angle $\alpha,$ as in Sect.\label{sect:two} describes $\overline{10}-8_{1}
$ mixing, angle $\phi$ is responsible for the mixing of Roper states with
already mixed $\overline{10}$, as depicted schematically in
Fig.\ref{fig:angles}.

Since we anticipate that the mixing will be not large, the physical states can
be labeled by the SU(3) representations so that:%
\begin{align}
\left\vert 8_{1}^{\text{phys}}\right\rangle  &  =\cos\alpha\left\vert
8_{1}\right\rangle +\sin\alpha
\left\vert \overline{10}\right\rangle ,\nonumber\\
\left\vert 8_{2}^{\text{phys}}\right\rangle  &  =-\sin\phi\sin\alpha\left\vert
8_{1}\right\rangle +\cos\phi\left\vert 8_{2}\right\rangle +\sin\phi\cos
\alpha\left\vert \overline{10}\right\rangle ,\nonumber\\
\left\vert \overline{10}^{\text{phys}}\right\rangle  &  =-\cos\phi\sin
\alpha\left\vert 8_{1}\right\rangle -\sin\phi\left\vert 8_{2}\right\rangle
+\cos\phi\cos\alpha\left\vert \overline{10}\right\rangle , \label{mixedstates}%
\end{align}

\subsection{Phenomenology of mixing}

\label{sect:phenomenology}

The sequential mixing (\ref{Omatrix}) allows to calculate physical masses by
iterative application of (\ref{physmass}). After the first mixing of
antidecuplet with the ground state octet, antidecuplet states satisfy
Gell-Man--Okubo mass formulae
(from now on particle names stand for their masses):%
\begin{equation}
N_{\overline{10}}=\Theta^{+}+\delta,\;\Sigma_{\overline{10}}=\Theta
^{+}+2\delta,\;\Xi_{\overline{10}}=\Theta^{+}+3\delta. \label{GO}%
\end{equation}
These states ($N_{\overline{10}}$ and  $\Sigma_{\overline{10}}$
then mix further with the Roper octet.%
\begin{align}
M_{8_{2}}  &  =M_{\overline{10}}^{\text{phys}}\sin^{2}\phi+M_{8_{2}%
}^{\text{phys}}\cos^{2}\phi,\nonumber\\
M_{\overline{10}}  &  =M_{\overline{10}}^{\text{phys}}\cos^{2}\phi+M_{8_{2}%
}^{\text{phys}}\sin^{2}\phi. \label{barmass}%
\end{align}
These formulae are valid both for nucleon- and sigma-like states. Since
physical masses of the nucleon-like states are known we can calculate bare
$N_{\overline{10}}$ from (\ref{barmass})%
\begin{equation}
N_{\overline{10}}=N_{\overline{10}}^{\text{phys}}\cos^{2}\phi+N_{8_{2}%
}^{\text{phys}}\sin^{2}\phi. \label{barmass1}%
\end{equation}
Since we know $\Theta
^{+}$ (which is not mixed) and $N_{\overline{10}}$ (from (\ref{barmass1})) we
can calculate $\delta$ and then $\Xi_{\overline{10}}^{\text{phys}}%
=\Xi_{\overline{10}}$ (because $\Xi_{\overline{10}}$ is not mixed as well) and
finally $\Sigma_{\overline{10}}^{\text{phys}}$ from the last equation in
(\ref{barmass}) since both $\Sigma_{8_{2}}^{\text{phys}}$ and $\Sigma
_{\overline{10}}$ (from (\ref{GO})) are known:%
\begin{align}
\Sigma_{\overline{10}}^{\text{phys}}  &  =\frac{1}{\cos^{2}\phi}\left(
2N_{\overline{10}}^{\text{phys}}\cos^{2}\phi+\left(  2N_{8_{2}}^{\text{phys}%
}-\Sigma_{8_{2}}^{\text{phys}}\right)  \sin^{2}\phi-\Theta^{+}\right)
,\label{Sigphys}\\
\Xi_{\overline{10}}^{\text{phys}}  &  =3\left(  N_{\overline{10}}%
^{\text{phys}}\cos^{2}\phi+N_{8_{2}}^{\text{phys}}\sin^{2}\phi\right)
-2\Theta^{+}. \label{Xiphys}%
\end{align}

\section{Decays in the presence of mixing}

\label{sect:decays}

To calculate the decay width of baryon $B_{1}\rightarrow B_{2}+\varphi$ (where
$\varphi$ stands for the pseudoscalar meson) we shall use -- following
\cite{dia,Ledwig:2008rw} -- the generalized Goldberger-Treiman relation
employed first by Witten, Adkins and Nappi in Ref.\cite{Adkins:1983ya}:
\begin{equation}
\Gamma_{B_{1}\rightarrow B_{2}\varphi}=\frac{g_{B_{1}B_{2}\varphi}^{2}}%
{2\pi(M_{1}+M_{2})^{2}}p_{\varphi}^{3}. \label{Gammadef}%
\end{equation}
Here $M_{1}$ is the mass of the decaying baryon, $M_{2}$ the mass of the decay
product, $p_{\varphi}$ meson momentum given by:%
\begin{equation}
p_{\varphi}=\frac{\sqrt{(M_{1}^{2}-(M_{2}+m_{\varphi})^{2})(M_{1}^{2}%
-(M_{2}-m_{\varphi})^{2})}}{2M_{1}}. \label{pphi}%
\end{equation}
The decay constant $g_{B_{1}B_{2}\varphi}$ stands for the matrix element of
the tensor decay operator $O_{\varphi}^{(8)}$:%
\begin{equation}
g_{B_{1}B_{2}\varphi}=\left\langle B_{2}^{\text{phys}}\right\vert O_{\varphi
}^{(8)}\left\vert B_{1}^{\text{phys}}\right\rangle . \label{gdef}%
\end{equation}
Explicit form of the decay operator is known for example in $\chi$QSM
\cite{dia}. Here we shall simply assume, that $O_{\varphi}^{(8)}$ transforms
as a $\varphi$ component of the octet and as spin $1$. This, together with the
assumption that $O_{\varphi}^{(8)}$ satisfies hermiticity condition will allow
us to express the relevant matrix elements between the unmixed states with the
help of SU$_{\text{fl}}$(3) isoscalar factors and the reduced matrix elements.
Below we list all matrix elements needed in the present analysis. Decays to
octet (ground state or Roper) are given by:%
\begin{align}
\left\langle 8,B_{2}\right\vert O_{\varphi}^{(8)}\left\vert \overline
{10},B_{1}\right\rangle  &  =-\left[
\begin{array}
[c]{cc}%
8 & 8\\
\varphi & B_{1}%
\end{array}
\right.  \left\vert
\begin{array}
[c]{c}%
\overline{10}\\
B_{2}%
\end{array}
\right]  G_{\overline{10}}\qquad\left(  \text{or }G_{\overline{10}}%
^{R}\right)  ,\nonumber\\
\left\langle 8,B_{2}\right\vert O_{\varphi}^{(8)}\left\vert 10,B_{1}%
\right\rangle  &  =\sqrt{2}\left[
\begin{array}
[c]{cc}%
8 & 8\\
\varphi & B_{1}%
\end{array}
\right.  \left\vert
\begin{array}
[c]{c}%
10\\
B_{2}%
\end{array}
\right]  G_{10}\qquad\left(  \text{or }G_{\overline{10}}^{R}\right)  .
\label{to8}%
\end{align}
Diagonal matrix elements are defined as follows:%
\begin{align}
\left\langle \overline{10},B_{2}\right\vert O_{\varphi}^{(8)}\left\vert
\overline{10},B_{1}\right\rangle  &  =\sqrt{2}\left[
\begin{array}
[c]{cc}%
8 & \overline{10}\\
\varphi & B_{1}%
\end{array}
\right.  \left\vert
\begin{array}
[c]{c}%
\overline{10}\\
B_{2}%
\end{array}
\right]  H_{\overline{10}},\nonumber\\
\left\langle 8,B_{2}\right\vert O_{\varphi}^{(8)}\left\vert 8,B_{1}%
\right\rangle  &  =2\left[
\begin{array}
[c]{cc}%
8 & 8\\
\varphi & B_{1}%
\end{array}
\right.  \left\vert
\begin{array}
[c]{c}%
8_{1}\\
B_{2}%
\end{array}
\right]  A+\sqrt{20}\left[
\begin{array}
[c]{cc}%
8 & 8\\
\varphi & B_{1}%
\end{array}
\right.  \left\vert
\begin{array}
[c]{c}%
8_{1}\\
B_{2}%
\end{array}
\right]  B, \label{diag}%
\end{align}
with $A,B\rightarrow A^{R},B^{R}$ when one of the octets is Roper. Note that
transitions $8\rightarrow\overline{10}$ are equal to $\overline{10}$
$\rightarrow8$ of (\ref{to8}) by the hermiticity requirement. For transitions
$8\rightarrow10$ and $10\rightarrow8$ more care is needed since decuplet has
spin 3/2; we shall comment upon this later. Matrix elements used in this paper
are displayed in Table \ref{tablematel}
\renewcommand{\arraystretch}{2}
\begin{table}[ptb]
\begin{tabular}
[c]{r|ccc}\hline
$B_{1}\rightarrow B_{2} \varphi$~ & ~ $\left\langle 8, B_{2}\right\vert
O_{\varphi}^{(8)}\left\vert \overline{10}, B_{1}\right\rangle $ &
$\left\langle \overline{10}, B_{2} \right\vert O_{\varphi}^{(8)}\left\vert
\overline{10}, B_{1}\right\rangle $ & $\left\langle 8, B_{2} \right\vert
O_{\varphi}^{(8)}\left\vert 8, B_{1} \right\rangle $\\\hline
$\Theta^{+}\rightarrow NK$ & $G_{\overline{10}}$ & $H_{\overline{10}}$ &
$-$\\\hline
$N\rightarrow N\pi~ $ & $\frac{1}{2}G_{\overline{10}}$ & $\frac{1}%
{2}H_{\overline{10}}$ & $A-3B$\\
$N\eta~ $ & $\frac{1}{2}G_{\overline{10}}$ & $-\frac{1}{2}H_{\overline{10}}$ &
$-A-B$\\
$K\Lambda~ $ & $-\frac{1}{2}G_{\overline{10}}$ & $-$ & $A-B$\\
$K\Sigma~ $ & $\frac{1}{2}G_{\overline{10}}$ & $H_{\overline{10}}$ &
$A+3B$\\\hline
$\Sigma\rightarrow N\overline{K}~$ & $\frac{1}{\sqrt{6}}G_{\overline{10}}$ &
$-\frac{2}{\sqrt{6}}H_{\overline{10}}$ & $-\frac{2}{\sqrt{6}}(A+3B)$\\
$\Sigma\eta~ $ & $\frac{1}{2}G_{\overline{10}}$ & $-$ & $2B$\\
$\Lambda\pi~ $ & $-\frac{1}{2}G_{\overline{10}}$ & $-$ & $2B$\\
$\Sigma\pi~ $ & $\frac{1}{\sqrt{6}}G_{\overline{10}}$ & $\frac{2}{\sqrt{6}%
}H_{\overline{10}}$ & $\frac{4}{\sqrt{6}}A$\\
$\Xi K$~ & $-\frac{1}{\sqrt{6}}G_{\overline{10}}$ & $-$ & $\frac{2}{\sqrt{6}%
}(A-3B)$\\\hline
$\Xi\rightarrow Xi\pi~$ & $-\frac{1}{\sqrt{2}} G_{\overline{10}}$ & $-$ &
$-$\\
$\Sigma K~ $ & $\frac{1}{\sqrt{2}} G_{\overline{10}}$ & $-\frac{1}{\sqrt{2}}
H_{\overline{10}}$ & $-$\\\hline
\end{tabular}
\caption{Matrix elements of the decay operator between SU$_{\mathrm{fl}}$(3)
symmetry states.}%
\label{tablematel}
\end{table}
\renewcommand{\arraystretch}{1}

In the case of pion-nucleon coupling and Roper decays we may safely neglect
small mixing corrections proportional to the sinuses of the mixing angles.
With notation of Table \ref{tablematel} we have:%
\begin{equation}
g_{\pi NN}=\frac{\cos^{2}\alpha}{\sqrt{3}}(A-3B),\qquad\varepsilon=\frac{F}%
{D}=-\frac{A}{3B}. \label{gpiNN}%
\end{equation}
Furthermore Roper decay constants read:%
\begin{align}
g_{RN\pi}  &  =\cos\phi\cos\alpha(A^{R}-3B^{R}),\nonumber\\
g_{RN\eta}  &  =-\cos\phi\cos\alpha(A^{R}+B^{R}). \label{Roper}%
\end{align}
Since for the Roper octet $\epsilon^R \approx 0.37$ \cite{guz2}, which
makes $g_{RN\eta}$ very small (note that the tiny decay width of
the Roper to $N\eta$ \cite{PDG} results both from the smallness
of the coupling constant and of the phase space), we will assume in
the following that%
\begin{equation}
B^{R}=-A^{R} \label{ABRoper}%
\end{equation}
which corresponds to%
\begin{equation}
\varepsilon^{R}=\frac{1}{3}. \label{epsRoper}%
\end{equation}

For antidecuplet decays we keep mixing terms, since they are comparable in
magnitude to the primary decay constants. For $\Theta^{+}$ we get:
\begin{equation}
g_{\theta NK}=\cos\alpha\,G_{\overline{10}}+\sin\alpha\,H_{\overline{10}}.
\label{gThNK}%
\end{equation}
Since $g_{\theta NK}$ can be directly read off from $\Theta^{+}$ decay width,
it is convenient to express the remaining decay constants through $g_{\theta
NK}$ and $g_{\pi NN}$, $\varepsilon$ and $g_{RN\pi}$. This leads to the
following decay constants for $N_{\overline{10}}$:%
\begin{align}
g_{N_{\overline{10}}N\pi}  &  =\frac{1}{2}\cos\phi\cos\alpha\,g_{\theta
NK}-\,\cos\phi\tan\alpha\sqrt{3}g_{\pi NN}-\tan\phi\,g_{RN\pi}\nonumber\\
&  \simeq\frac{1}{2}g_{\theta NK}-\sin\alpha\sqrt{3}g_{\pi NN}-\sin
\phi\,g_{RN\pi},\nonumber\\
g_{N_{\overline{10}}N\eta}  &  =\frac{1}{2}\cos\phi\cos\alpha\,g_{\theta
NK}-\frac{1}{2}\cos\phi\sin2\alpha\,H_{\overline{10}}+\cos\phi\tan\alpha
\frac{3\varepsilon-1}{1+\varepsilon}\frac{g_{\pi NN}}{\sqrt{3}}\nonumber\\
&  \simeq\frac{1}{2}\,g_{\theta NK}-\sin\alpha\,H_{\overline{10}}+\sin
\alpha\,\frac{3\varepsilon-1}{1+\varepsilon}\frac{g_{\pi NN}}{\sqrt{3}}.
\label{gN10bar}%
\end{align}
Approximate equalities in Eq.(\ref{gN10bar}) correspond to the small mixing
angle limit ($\cos($angle$)=1$).

Apart from physical decay constants mentioned above, $g_{N_{\overline{10}%
}N\eta}$ depends additionally on $H_{\overline{10}}$ that canceled in the
expression for $g_{N_{\overline{10}}N\pi}$. This is \emph{a priori}, apart
from the mixing angles, new free parameter that cannot be constrained from the
data. In what follows we shall estimate $H_{\overline{10}}$ using model
calculations within the framework of $\chi$QSM.

Decay constant to $\Lambda K$ reads%
\begin{align}
g_{N_{\overline{10}}\Lambda K}  &  =-\frac{1}{2}\cos\phi g_{\theta NK}%
+\frac{1}{2}\cos\phi\sin\alpha\,H_{\overline{10}}\nonumber\\
&  -\frac{1}{2}\frac{\tan\phi}{\cos\alpha}g_{RN\pi}-\frac{\cos\phi\tan\alpha
}{\cos\alpha}\frac{3\varepsilon+1}{1+\varepsilon}\frac{g_{\pi NN}}{\sqrt{3}%
}\nonumber\\
&  \simeq-\frac{1}{2}\left(  g_{\theta NK}-\sin\alpha\,H_{\overline{10}}%
+\sin\phi g_{RN\pi}+\sin\alpha\frac{3\varepsilon+1}{1+\varepsilon}%
\frac{2g_{\pi NN}}{\sqrt{3}}\right)  . \label{g10bar2}%
\end{align}

In the SU$_{\text{fl}}$(3) limit antidecuplet states cannot decay to decuplet.
However, in the presence of mixing such decays are possible:%
\begin{align}
g_{N_{\overline{10}}\Delta\pi}  &  =-2\cos\phi\tan\alpha\,g_{\Delta N\pi}%
-\tan\phi\,g_{R\Delta\pi}\nonumber\\
&  \simeq-2\sin\alpha\,g_{\Delta N\pi}-\sin\phi\,g_{R\Delta\pi}
\label{g10bar10}%
\end{align}
where we have introduced new decay constant $g_{R\Delta\pi}$ describing Roper
decay to $\Delta\pi$. Here a remark concerning factor 2 in front of $g_{\Delta
N\pi}$ in Eq.(\ref{g10bar10}) is in order. Factor 2 implies%
\begin{equation}
g_{N\Delta\pi}=2g_{\Delta N\pi}. \label{gDelN}%
\end{equation}
This relation follows from the fact, that the decay operator transforms as an
octet in SU$_{\text{fl}}$(3) but has also spin 1, since the decays considered
here occur in $P$-wave. While calculating the width we average over initial
spin and isospin and sum over the final state. In the case of $\Delta$ decay
initial spin and isospin are $3/2$ and averaging of amplitude $A_{\Delta
\rightarrow N\pi}$ gives%
\begin{equation}
\left\langle A_{\Delta\rightarrow N\pi}^{2}\right\rangle =\frac{1}{2S^{\Delta
}+1}\frac{1}{2T^{\Delta}+1}%
{\displaystyle\sum\limits_{S_{z}^{\Delta},S_{z}^{N}}}
{\displaystyle\sum\limits_{T_{z},T_{z}^{\Delta},T_{z}^{N}}}
A^{2}=\frac{1}{16}%
{\displaystyle\sum\limits_{S_{z}^{\Delta},S_{z}^{N}}}
{\displaystyle\sum\limits_{T_{z},T_{z}^{\Delta},T_{z}^{N}}}
A^{2},
\end{equation}
where $A$ is the reduced amplitude. When calculating transition $N\rightarrow\Delta\pi$ which appears due to the mixing
$N_{\overline{10}}-N_{8_1}$
we have, due to hermiticity, the same amplitude $A$,
however averaging is
different:%
\begin{equation}
\left\langle A_{N\rightarrow\Delta\pi}^{2}\right\rangle =\frac{1}{2S^{N}%
+1}\frac{1}{2T^{N}+1}%
{\displaystyle\sum\limits_{S_{z}^{\Delta},S_{z}^{N}}}
{\displaystyle\sum\limits_{T_{z},T_{z}^{\Delta},T_{z}^{N}}}
A^{2}=\frac{1}{4}%
{\displaystyle\sum\limits_{S_{z}^{\Delta},S_{z}^{N}}}
{\displaystyle\sum\limits_{T_{z},T_{z}^{\Delta},T_{z}^{N}}}
A^{2}.
\end{equation}
Hence%
\begin{equation}
\sqrt{\left\langle A_{N\rightarrow\Delta\pi}^{2}\right\rangle }=2\sqrt
{\left\langle A_{\Delta\rightarrow N\pi}^{2}\right\rangle }%
\end{equation}
which is effectively (up to the overall phase) the same as (\ref{gDelN}).

Decay constant for $\Sigma_{\overline{10}}$ to nucleon reads as follows:%
\begin{align}
g_{\Sigma_{\overline{10}}N\overline{K}}  &  =\frac{1}{\sqrt{6}}\left(
\cos\phi\cos\alpha\,g_{\theta NK}-\frac{3}{2}\cos\phi\sin2\alpha
\,H_{\overline{10}}\right. \nonumber\\
&  \left.  -\tan\phi\,g_{RN\pi}-\cos\phi\tan\alpha\frac{1-\varepsilon
}{1+\varepsilon}2\sqrt{3}g_{\pi NN}\right) \nonumber\\
&  \simeq\frac{1}{\sqrt{6}}\left(  g_{\theta NK}-3\sin\alpha\,H_{\overline
{10}}-\sin\phi g_{RN\pi}-\sin\alpha\frac{1-\varepsilon}{1+\varepsilon}%
2\sqrt{3}g_{\pi NN}\right)  .
\end{align}
Decays to $\Sigma$ take the following form:%
\begin{align}
g_{\Sigma_{\overline{10}}\Sigma\pi}  &  =\frac{1}{\sqrt{6}}\left(  \cos
\phi\cos\alpha\,g_{\theta NK}+\frac{1}{2}\cos\vartheta\cos\phi\sin
2\alpha\,H_{\overline{10}}\right. \nonumber\\
&  \left.  -\tan\phi\,g_{RN\pi}-\cos\phi\tan\alpha\frac{4\varepsilon
}{1+\varepsilon}\sqrt{3}g_{\pi NN}\right) \nonumber\\
&  \simeq\frac{1}{\sqrt{6}}\left(  g_{\theta NK}+\sin\alpha\,H_{\overline{10}%
}-\sin\phi g_{RN\pi}-\sin\alpha\frac{4\varepsilon}{1+\varepsilon}\sqrt
{3}g_{\pi NN}\right) \\
g_{\Sigma_{\overline{10}}\Sigma\eta}  &  =\frac{1}{2}\left(  \cos\phi
\cos\alpha\,g_{\theta NK}-\frac{1}{2}\cos\phi\sin2\alpha\,H_{\overline{10}%
}\right. \nonumber\\
&  \left.  +\tan\phi\,g_{RN\pi}+\cos\phi\tan\alpha\frac{4}{1+\varepsilon}%
\frac{g_{\pi NN}}{\sqrt{3}}\right) \nonumber\\
&  \simeq\frac{1}{2}\left(  g_{\theta NK}-\sin\alpha\,H_{\overline{10}}%
+\sin\phi\,g_{RN\pi}+\sin\alpha\frac{4}{1+\varepsilon}\frac{g_{\pi NN}}%
{\sqrt{3}}\right)  .
\end{align}
Decay to $\Lambda$ is given by%
\begin{align}
g_{\Sigma_{\overline{10}}\Lambda\pi}  &  =\frac{1}{2}\left(  -\cos
\phi\,g_{\theta NK}+\frac{{}}{{}}\cos\phi\sin\alpha\,H_{\overline{10}}\right.
\nonumber\\
&  \left.  +\frac{\tan\phi}{\cos\alpha}g_{RN\pi}+\frac{\cos\phi\tan\alpha
}{\cos\alpha}\frac{4\varepsilon}{1+\varepsilon}\sqrt{3}g_{\pi NN}\right)
\nonumber\\
&  \simeq-\frac{1}{2}\left(  g_{\theta NK}-\sin\alpha\,H_{\overline{10}}%
-\sin\phi\,g_{RN\pi}-\sin\alpha\frac{4\varepsilon}{1+\varepsilon}\sqrt
{3}g_{\pi NN}\right)  .
\end{align}
Finally for the decays to decuplet we have:%
\begin{align}
g_{\Sigma_{\overline{10}}\Sigma^{\ast}\pi}  &  =-\frac{1}{\sqrt{6}%
}g_{N\overline{_{10}}\Delta\pi},\nonumber\\
g_{\Sigma_{\overline{10}}\Delta\pi}  &  =\sqrt{\frac{2}{3}}g_{N\overline
{_{10}}\Delta\pi}.
\end{align}

\section{Where are and what are properties of strange particles in
antidecuplet}

\label{sect:where}

In order to get some numerical insight into Eqs.(\ref{gN10bar}) let us recall
the results of $\chi$QSM%
\begin{equation}
G_{\overline{10}}=\sqrt{\frac{3}{5}}G_{\overline{10}}^{\chi\text{QSM}},\quad
H_{\overline{10}}=\frac{\sqrt{3}}{4}H_{\overline{10}}^{\chi\text{QSM}}%
\end{equation}
in notation of Ref.\cite{michal1}. Taking fits to decays without mixing we get%
\begin{equation}
G_{\overline{10}}\sim1.3,\quad H_{\overline{10}}\sim-6.9.
\end{equation}
In the present work we completely eliminate $G_{\overline{10}}$ through
Eq.(\ref{gThNK}), however $H_{\overline{10}}$ reappears in other decay
constants. Therefore in the following we shall choose $H_{\overline{10}}=-7$,
however we also checked other values, namely $-5$ and $-9$. Note that even if
we know some decay constant -- such as $g_{\pi NN}$ -- from experiment, we
still have freedom in choosing the relative phase with which it enters
Eqs.(\ref{gThNK})--(\ref{g10bar10}). Some of these phases can be absorbed to
the sign of the mixing angles, the other ones have to be chosen arbitrarily.
For example taking $g_{\pi NN}>0$ we can absorb the relative phase into the
sign of $\alpha$ (see Eq.(\ref{gThNK})). Therefore the sign of $H_{\overline
{10}}$ cannot be absorbed into redefinition of $\alpha$. Choosing negative
$H_{\overline{10}}$ we have followed the sign dictated by the $\chi$QSM,
therefore we have to check sensitivity of our predictions for $H_{\overline
{10}}>0$. We find that in this case the results are not compatible with
experimentally acceptable pattern of decay widths. Similarly the phase of
$g_{RN\pi}$ can be absorbed into the sign of angle $\phi$, therefore the
relative phase between $g_{\Delta N\pi}$ and $g_{R\Delta\pi}$ is not fixed
(see Eq.(\ref{g10bar10})). In what follows we assume that both $g_{\Delta
N\pi}$ and $g_{R\Delta\pi}$ are positive and show that for negative phase we
again get results that ar not compatible with experiment.

Throughout this paper we assume the following values for the parameters
entering decay constants:%
\begin{equation}
g_{\pi NN}=13.21,\quad\varepsilon=0.56,\quad\varepsilon^{R}=\frac{1}{3}%
\end{equation}
and the following decay widths for $\Delta$ and Roper \cite{PDG}:%
\begin{equation}
\Gamma_{\Delta\rightarrow N\pi}=120\;\text{MeV},\quad\Gamma_{R\rightarrow
N\pi}=152.1\;\text{MeV},\quad\Gamma_{R\rightarrow\Delta\pi}=58.5\;\text{MeV}%
\end{equation}

\renewcommand{\baselinestretch}{0.5}

\begin{figure}[h]
\begin{centering}
\includegraphics[width=6.0cm]{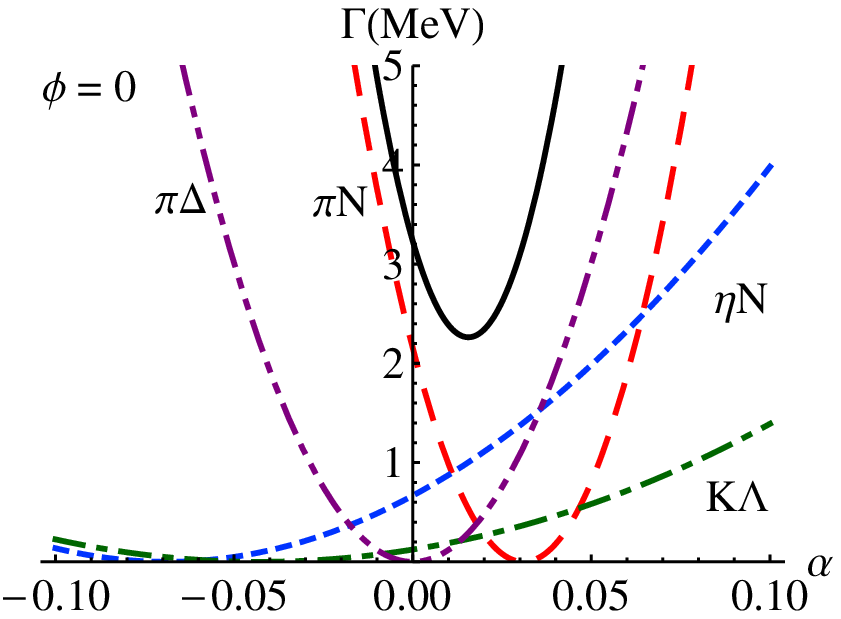}~~
\includegraphics[width=6.0cm]{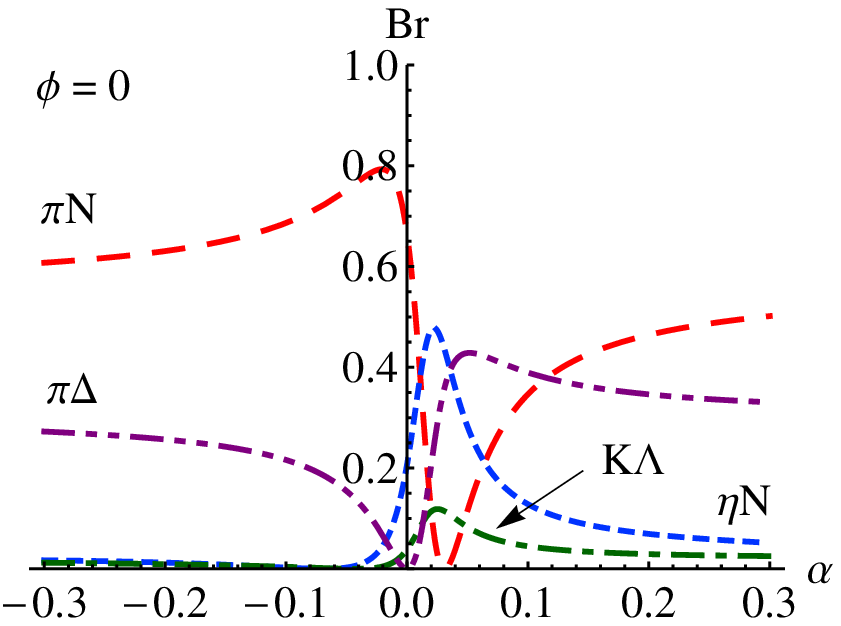}\\
\includegraphics[width=6.0cm]{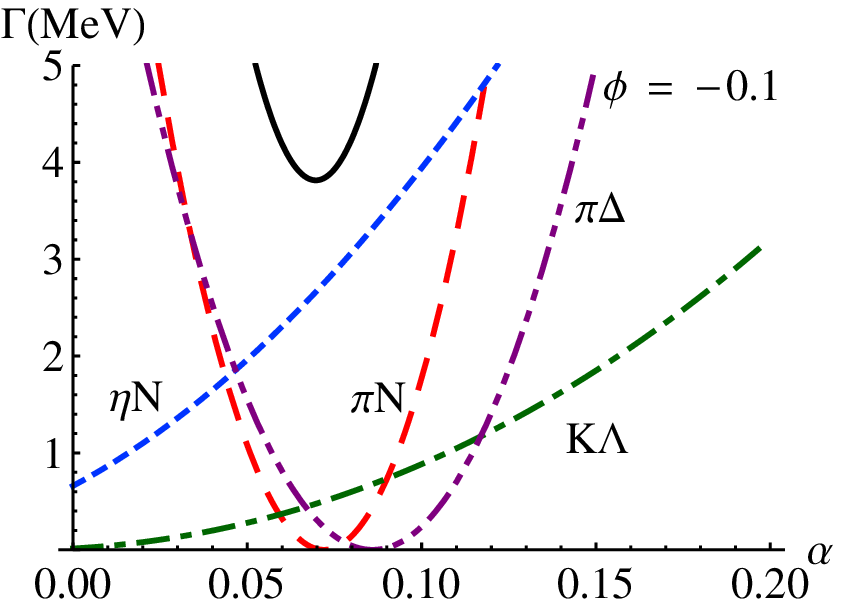}~~
\includegraphics[width=6.0cm]{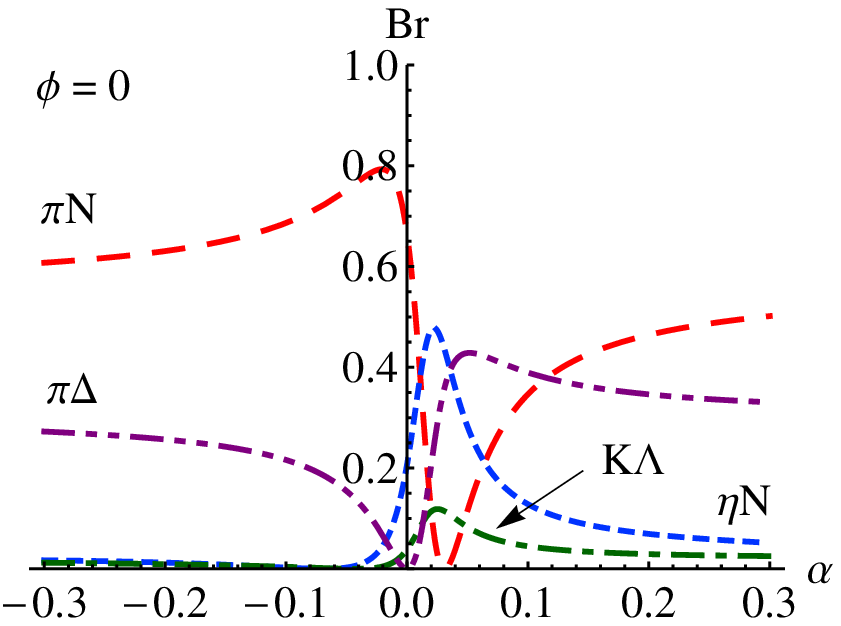}\\
\par\end{centering}
\caption{Partial widths an branching ratio for $N_{\overline{10}}$ decays
into: $N\pi$ -- long dash (red), $N\eta$ -- short dash (blue), $\Lambda K$ --
dash-dot (dark green), $\Delta\pi$ -- dash-dot-dot (purple) and total width --
solid (black) as functions of angle $\alpha$ for fixed $\phi$: $\phi=0$ --
upper plots, $\phi=-0.1$ -- lower plots. For all plots $H_{\overline{10}}=-7$.
}%
\label{fig:hm7al}%
\end{figure}
\renewcommand{\baselinestretch}{1.0}

\renewcommand{\baselinestretch}{0.5}
\begin{figure}[h]
\begin{centering}
\includegraphics[width=6.0cm]{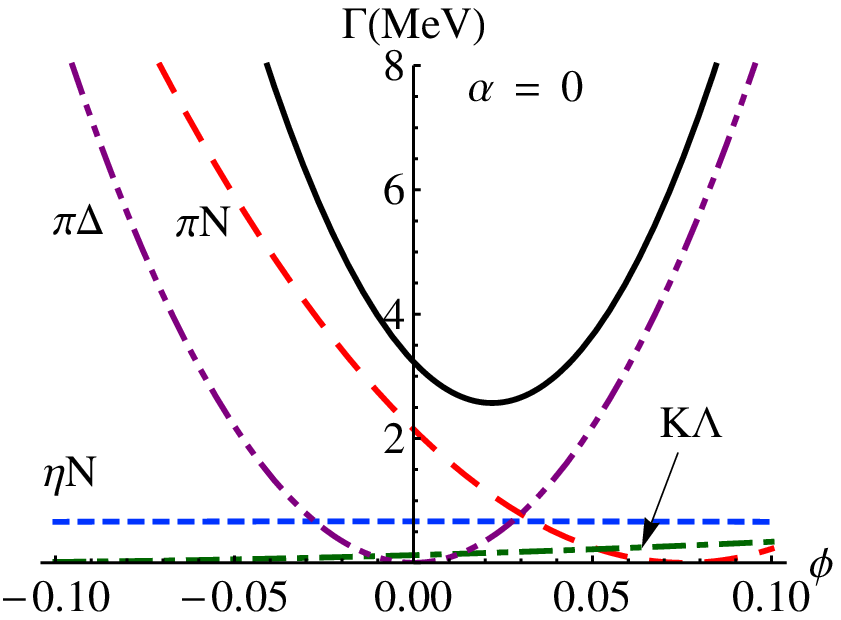}~~
\includegraphics[width=6.0cm]{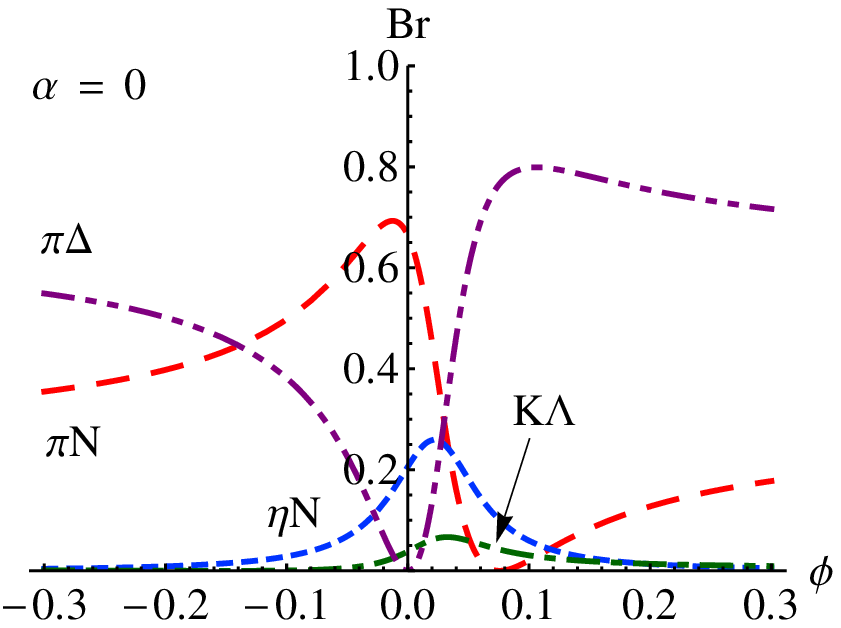}\\
\includegraphics[width=6.0cm]{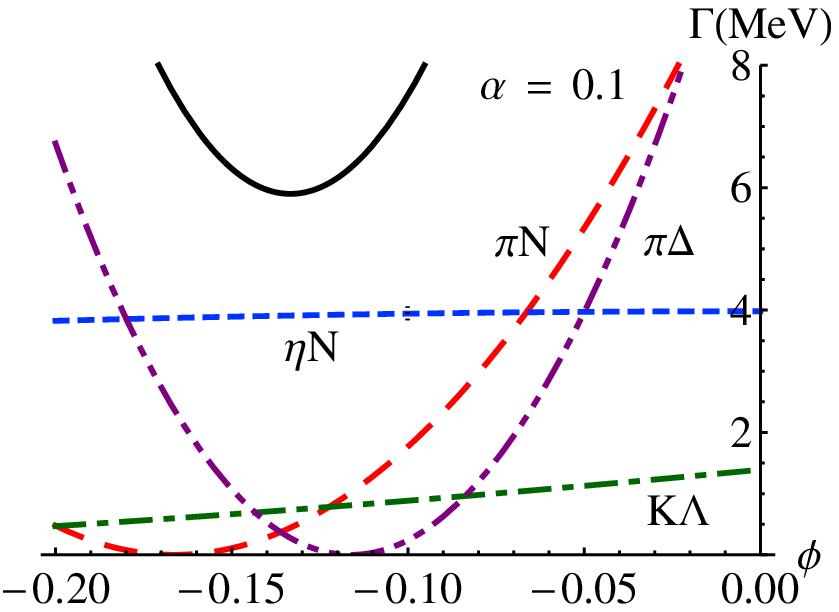}~~
\includegraphics[width=6.0cm]{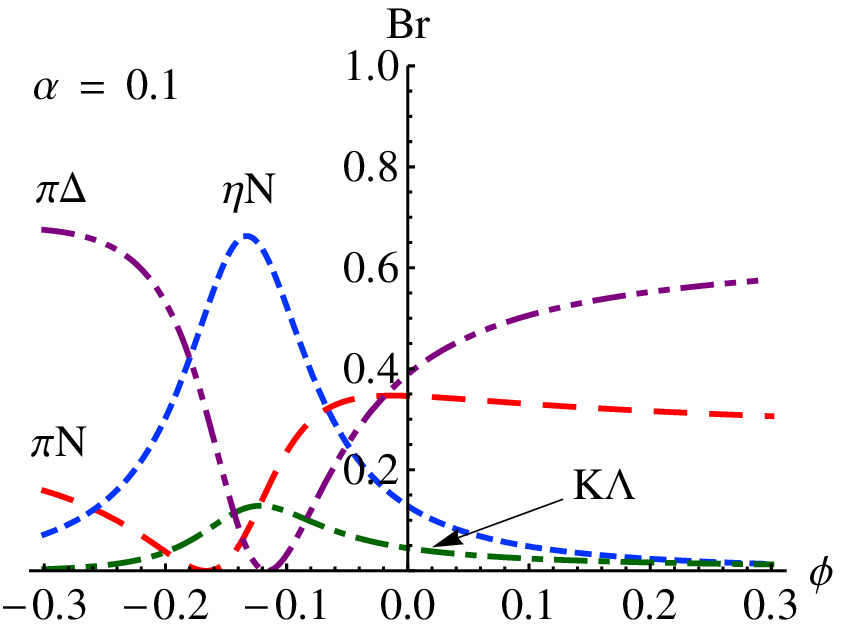}\\
\par\end{centering}
\caption{Same as Fig. \ref{fig:hm7al} but as functions of angle $\phi$ for
fixed $\alpha$: $\alpha=0$ -- upper plots, $\alpha=+0.1$ -- lower plots.}%
\label{fig:hm7fi}%
\end{figure}

\renewcommand{\baselinestretch}{1}

\begin{figure}[h]
\begin{centering}
\includegraphics[scale=0.8]{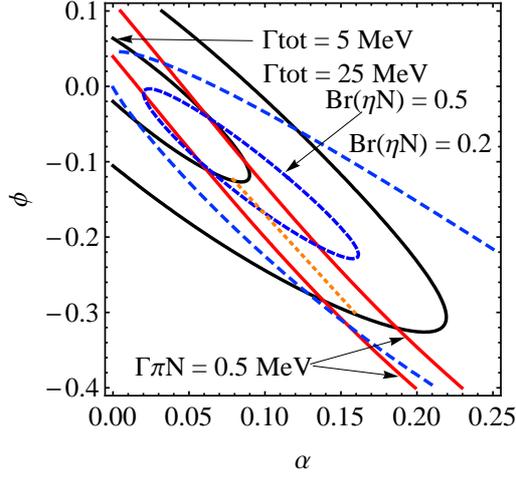}
\par\end{centering}
\caption{Contour plot corresponding to the conditions (\ref{conditions}). We
also plot an orange dotted line inside the allowed region of mixing angles
along which we later plot partial decays widths and branching ratios.}%
\label{fig:contour}%
\end{figure}

In Figs. \ref{fig:hm7al} and \ref{fig:hm7fi} we plot partial decay widths and
branching ratios for $N_{\overline{10}}$ decays as functions of mixing angles
$\alpha$ and $\phi$. Total width has been calculated by adding all partial
widths plus 10\% for unaccounted three body decays. First of all let us
observe that without mixing the dominant decay mode $\pi N$ has branching
ratio of 66\%, next is $\eta N$ with branching ratio 20\% and finally
$K\Lambda$ -- 4\%. Decays to decuplet are forbidden. This decay pattern
contradicts experiment (provided we want to interpret $N(1685)$ which decays
predominantly to $\eta N$, as a cryptoexotic member of antidecuplet). However,
the branching ratios and decay widths change rather rapidly when mixing is
included. We can see from Figs. \ref{fig:hm7al} and \ref{fig:hm7fi} that for
positive $\alpha$ and negative $\phi$ there exist regions where the decay to
$\eta N$ is dominant and the total width is relatively large. In order to find
the preferable region in $(\alpha,\phi)$ plane we impose the following
requirements \cite{str,Kuznetsov:2008hj,Kuznetsov:2008ii}:%
\begin{align}
\Gamma(N_{\overline{10}}  &  \rightarrow\pi N)<0.5\;\text{MeV,}\nonumber\\
\text{Br}(N_{\overline{10}}  &  \rightarrow\eta N)>0.2,\nonumber\\
5\;\text{MeV}  &  <\Gamma_{\text{tot}}(N_{\overline{10}})<25\;\text{MeV}
\label{conditions}%
\end{align}
and plot pertinent contours in Fig.\ref{fig:contour}. The allowed region
defined in Eq.(\ref{conditions}) should lie between the two continuous (black)
ellipses corresponding to the allowed range of the total width, inside the
outer dashed blue ellipse corresponding to Br$(N_{\overline{10}}%
\rightarrow\eta N)=0.2$, and between two solid (red) lines where
$\Gamma(N_{\overline{10}}\rightarrow\pi N)<5\;$MeV. We see that there is a
rather narrow strip of allowed angles concentrated in the vicinity of the
point $\alpha\approx0.11$ and $\phi\approx-0.2$. In Fig. \ref{fig:contour} we
also plot the dotted (orange) line inside the allowed region given by the
equation
\begin{equation}
\phi(\alpha)=0.0508-2.207\alpha,\quad0.079<\alpha<0.159. \label{or_lin}%
\end{equation}

\renewcommand{\baselinestretch}{0.5} \renewcommand{\baselinestretch}{1}

\begin{figure}[h]
\begin{centering}
\includegraphics[width=6.0cm]{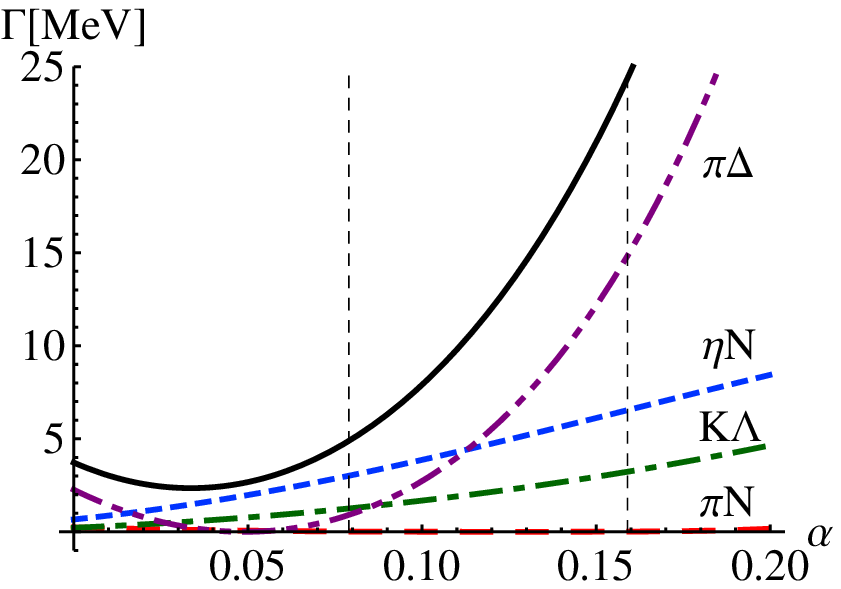}~~
\includegraphics[width=6.0cm]{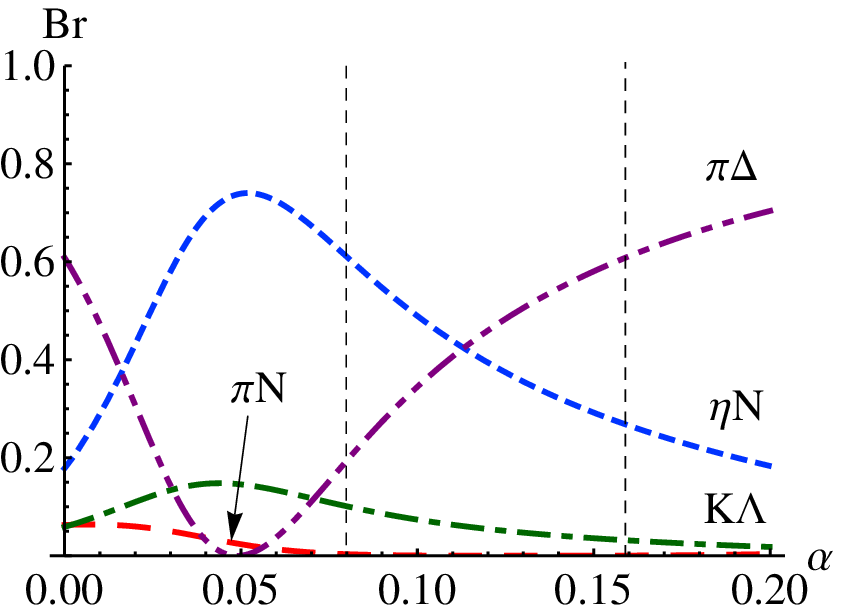}
\par\end{centering}
\caption{Partial widths an branching ratio for $N_{\overline{10}}$ decays
into: $N\pi$ -- long dash (red), $N\eta$ -- short dash (blue), $\Lambda K$ --
dash-dot (dark green), $\Delta\pi$ -- dash-dot-dot (purple) and total width --
solid (black) plotted along the line (\ref{or_lin}). For all plots
$H_{\overline{10}}=-7$. }%
\label{fig:orange_line}%
\end{figure}

In the following figures we plot various quantities along the line
(\ref{or_lin}), \emph{i.e.} for mixing angles inside the allowed region,
indicating the limits on angle $\alpha$ by vertical thin lines. First in Fig.
\ref{fig:orange_line} we show partial decay widths and branching ratios of
$N_{\overline{10}}$. We see that indeed the decay to $\pi N$ is strongly
suppressed, however another channel, namely the decay to $\pi\Delta$, starts
to dominate for larger mixing angles. In any case decays to decuplet are large
(remember they are forbidden if there is no mixing) and this prediction
provides a stringent test of our model.

\renewcommand{\baselinestretch}{0.5} \renewcommand{\baselinestretch}{1}

\renewcommand{\baselinestretch}{0.5} \begin{figure}[h]
\begin{centering}
\includegraphics[scale=0.8]{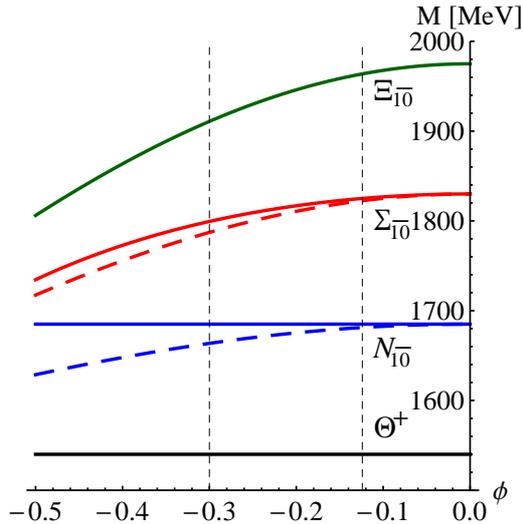}
\par\end{centering}
\caption{Masses of antidecuplet states as functions of angle $\phi$. Solid
lines correspond to the physical masses, whereas dashed lines to the GMO
masses before mixing with Roper octet. Note that for $\Theta^{+}$ and
$\Xi_{\overline{10}}$ GMO states and physical states coincide.}%
\label{fig:SigmaXi}%
\end{figure}\renewcommand{\baselinestretch}{1}

Next we come to the predictions for $\Sigma_{\overline{10}}$ and also for
$\Xi_{\overline{10}}$. In Fig. \ref{fig:SigmaXi} we plot masses of
antidecuplet states as functions of $\phi$. Solid lines correspond to the
physical masses. Since we take $\Theta^{+}$ and $N_{\overline{10}}$ as input
they do not depend on mixing angle. Dashed lines correspond to GMO states
before mixing. We see that%
\begin{align}
1795\;\text{MeV}  &  <M_{\Sigma_{\overline{10}}}<1830\;\text{MeV,}%
\label{sigmarange}\\
1900\;\text{MeV}  &  <M_{\Xi_{\overline{10}}}<1970\;\text{MeV} \label{xirange}%
\end{align}
within the allowed limits of Fig. \ref{fig:contour} (not only at the endpoints
of the line (\ref{or_lin})).

\begin{figure}[h]
\begin{centering}
\includegraphics[width=6.0cm]{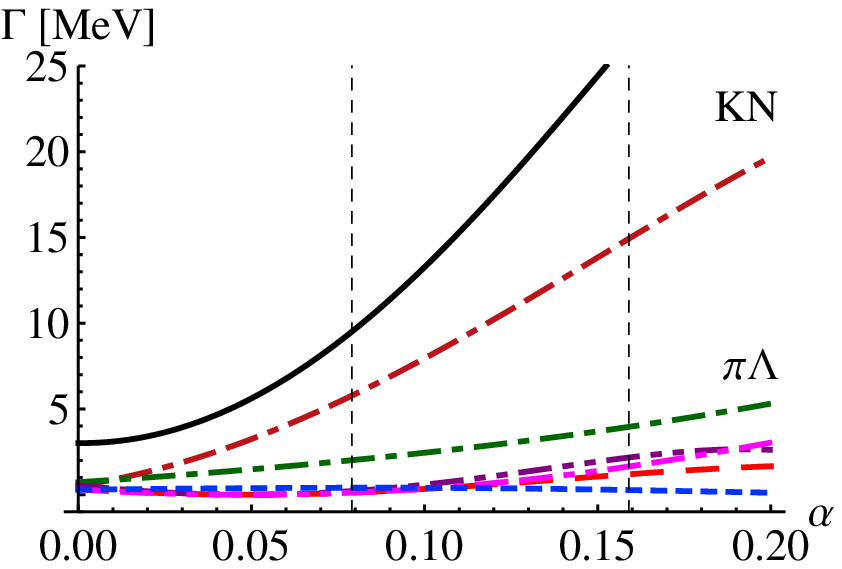}~~
\includegraphics[width=6.0cm]{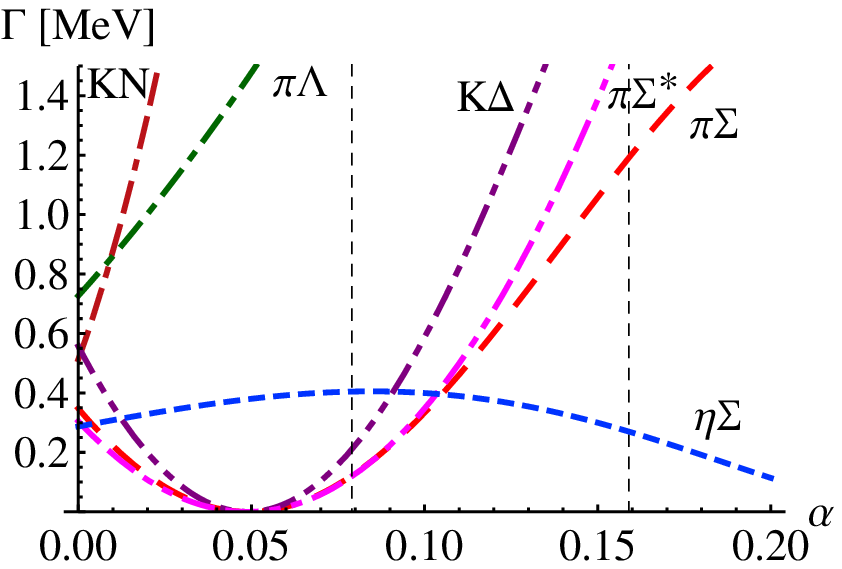}
\par\end{centering}
\caption{Partial widths for $\Sigma_{\overline{10}}$ decays into: NK -- double
dash dot (brown), $\Lambda\pi$ -- dash dot (dark green), $\Sigma\pi$ -- long
dash (red), $\Delta$K -- double dash dot (purple), $\Sigma\eta$ -- short dash
(blue), $\Sigma^{\ast}\pi$ -- dash-dot-dot (pink) and total width -- solid
(black) plotted along the line (\ref{or_lin}). For all plots $H_{\overline
{10}}=-7$. On right panel we present the enlargement of the left plot in order
to distinguish different decay modes that are below 1.2 MeV}%
\label{fig:GSig_orange_line}%
\end{figure}

\begin{figure}[h]
\begin{centering}
\includegraphics[width=6.0cm]{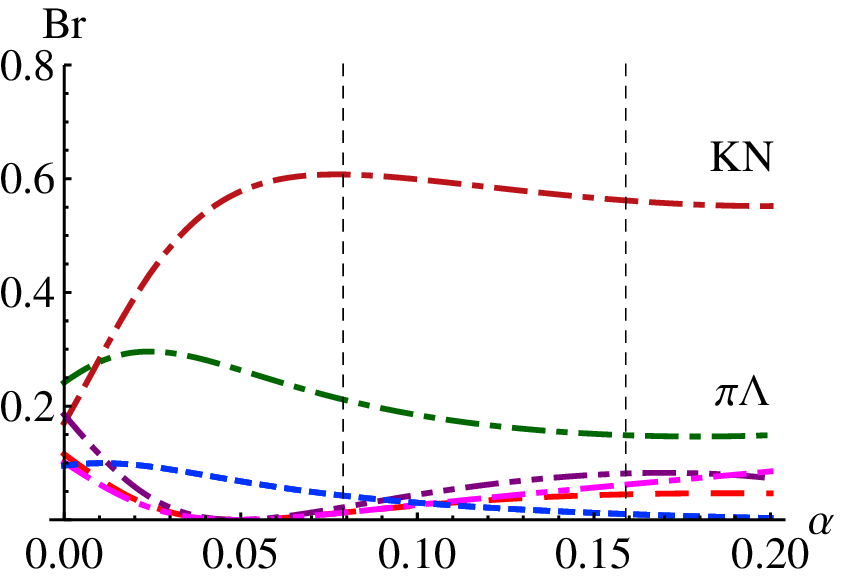}~~
\includegraphics[width=6.0cm]{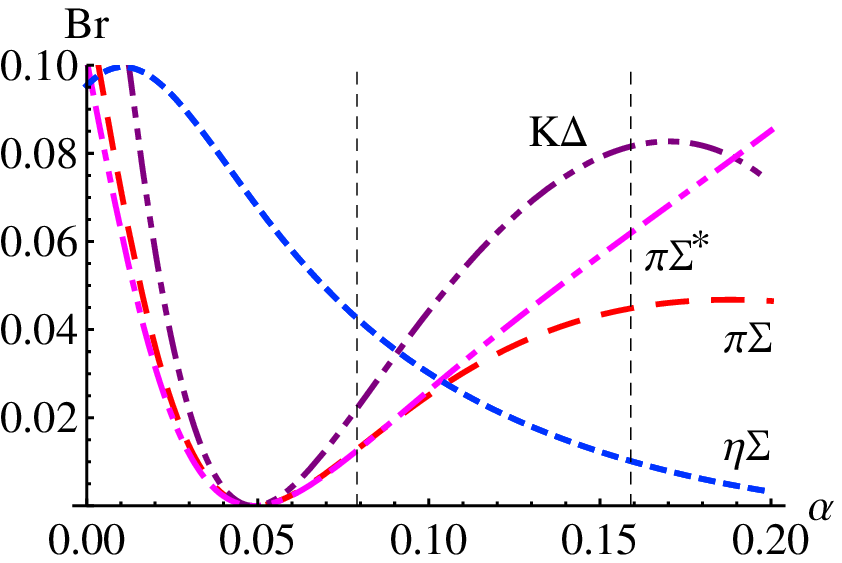}
\par\end{centering}
\caption{Branching ratios for decays of $\Sigma_{\overline{10}}$. All lines as
in Fig. \ref{fig:GSig_orange_line}. }%
\label{fig:BrSig_orange_line}%
\end{figure}

Finally in Fig. \ref{fig:GSig_orange_line} we plot partial widths for the
decays of $\Sigma_{\overline{10}}$. We see two dominating decay modes:
$\Sigma_{\overline{10}}\rightarrow KN$ and $\pi\Lambda$. In the right panel we
magnify the scale to distinguish between the remaining decays. In Fig.
\ref{fig:BrSig_orange_line} we plot the pertinent branching ratios. In Table
\ref{tablewidths}  we give the range of widths within the allowed limits of
Fig. \ref{fig:contour} (not only along the line (\ref{or_lin})).

\begin{table}[ptb]
\begin{tabular}
[c]{ccc}\hline
\text{mode} & $\Gamma_{\min}\text{[MeV]}$ & $\Gamma_{\max}\text{[MeV]}%
$\\\hline
$KN$ & 5.50 & 15.27\\
$\pi\Lambda$ & 1.70 & 4.40\\
$K\Delta$ & 0.09 & 2.57\\
$\pi\Sigma$ & 0.01 & 1.74\\
$\pi\Sigma^{\ast}$ & 0.05 & 1.95\\
$\eta\Sigma$ & 0.22 & 0.50\\\hline
$\text{total}$ & 9.7 & 26.9\\\hline
\end{tabular}
\caption{Range of decay widths of $\Sigma_{\overline{10}}$ in the region of
allowed mixing angles.}%
\label{tablewidths}
\end{table}

\renewcommand{\baselinestretch}{0.5} \renewcommand{\baselinestretch}{1}

\renewcommand{\baselinestretch}{0.5} \renewcommand{\baselinestretch}{1}

\renewcommand{\baselinestretch}{0.5} \begin{figure}[h]
\begin{centering}
\includegraphics[width=6.0cm]{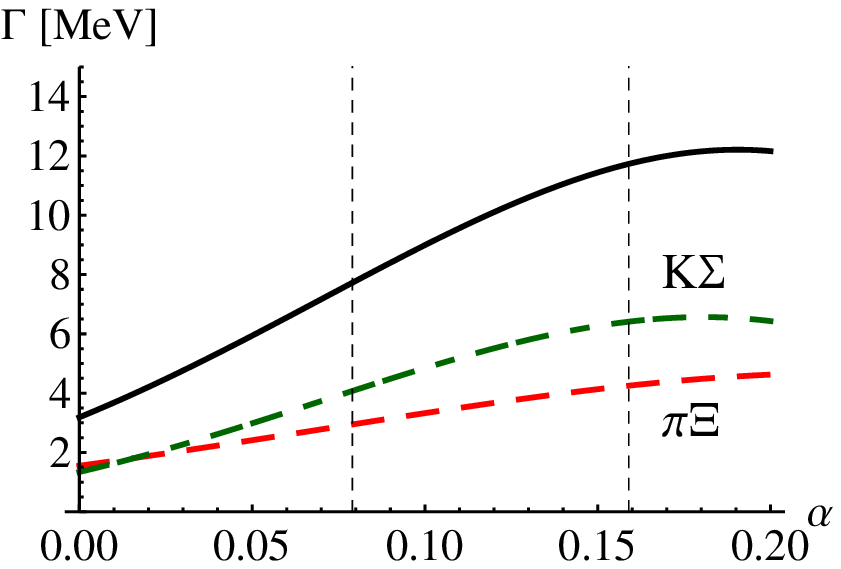}
\par\end{centering}
\caption{Partial widths for $\Xi_{\overline{10}}$ decays into: $N\pi$ -- long
dash (red), and total width -- solid (black) plotted along the line
(\ref{or_lin}). For all plots $H_{\overline{10}}=-7$.}%
\label{fig:GXi_orange_line}%
\end{figure}\renewcommand{\baselinestretch}{1}

Finally we also plot in Fig.\ref{fig:GXi_orange_line} partial decay widths for
the two allowed decays modes of $\Xi_{\overline{10}}$.

\section{Conclusions}

\label{sect:conclusions}

In the present paper we have examined mixing scenario in which the
crypts-exotic antidecuplet states mix with the respective states in the ground
state octet and in the Roper octet. This scenario is motivated by recent
experimental results on new narrow nucleon resonance $N(1685)$. Its small
decay width $\Gamma<25$ MeV and the fact that $n(1685)$ (neutron-like state)
undergoes photoexcitation in $\eta$ meson production and $p(1685)$ does not,
is easily explained if these states are interpreted as members of
antidecuplet. However $N(1685)$ seems to have very small coupling to the $\pi
N$ channel and comparatively large one to $\eta N$ channel in contradiction to
pure SU$_{\text{fl}}$(3) predictions for these decays. Since Roper resonance
has very small partial width to $\eta N$ one can adjust its mixing angle with
$N(1685)$ to suppress $\pi N$ coupling not affecting the $\eta$N one. We have
shown that there exists a small, but stable against variation of the unknown
couplings, region in the space of mixing angles where such mechanism is
effectively working. Our main result concerning the strange members of
antidecuplet is based on the observation that $\Sigma$-like states have the
same mixing angles as the nucleonic states and on the fact that $\Xi
_{\overline{10}}$ does not mix neither with Roper octet nor with the ground
state octet. This allows us to constrain masses of these states and their
decay patterns.

Before we discuss our results in more detail we want to stress that the mass
limits given in Eqs.(\ref{sigmarange},\ref{xirange}) and the decay widths
summarized in Tab. \ref{tablewidths} should be considered as the qualitative
ones. Variations of Roper decay constants within the experimental
limits and Roper $\epsilon_R$ parameter would enlarge limits derived in this
paper. Also our discussion concerning  $H_{\overline{10}}$ parameter
was only qualitative.

We predict that $\Sigma_{\overline{10}}$ has a mass around 1815 MeV within the
limits of Eq.(\ref{sigmarange}). There are no known $\Sigma$ resonances in
this energy range \cite{PDG}. Total width of $\Sigma_{\overline{10}}$ does not
exceed 30 MeV but is also constrained from below being larger that 10 MeV.
Note that we have calculated total widths by
adding all partial widths and 10\% for unaccounted  three body decays.
Most prominent decay channels are $KN$ and $\pi\Lambda$ with branching ratios
approximately 60\% and 20\% respectively. Due to the mixing SU$_{\text{fl}}%
$(3) forbidden decays to decuplet are possible, but small, at the level of 5
to 9\%.

The simplest way to detect $\Sigma_{\overline{10}}$ is its formation
in $K^- p$ scattering with the kaon beam of $p_{\rm lab}\sim 1$~GeV.
The corresponding resonance cross section is about 5~millibarn. Note, however, that
the non-resonant $K^- p$ cross-section  in this energy region is tens of millibarn.
That requires high statistics experiment and detailed partial wave analysis.
Additionally, the small width of predicted $\Sigma_{\overline{10}}$ demands high energy resolution
of the kaon beam. $\Sigma_{\overline{10}}$ can be produced in the photoproduction experiments, e.g.
in $\gamma+ p\to \Sigma_{\overline{10}}+K_S$, note however that according to simple estimate
based on the U-spin the corresponding production cross section is about three times smaller than
the analogous $\gamma+ p\to \Theta^++K_S$ cross section, i.e. very small. One can reveal
the small signal of the $\gamma+ p\to \Sigma_{\overline{10}}+K_S\to p+K_L+K_S$ processes using its interference
with much stronger amplitude of $\gamma+p\to \phi +p\to p+K_L+K_S$ \cite{interf}. Details of possible ways to see
$\Sigma_{\overline{10}}$ in various processes we shall give elsewhere \cite{elsewhere}.

From (\ref{xirange}) we see that $M_{\Xi_{\overline{10}}}$ is larger than 1900
MeV. The latter estimate is in disagreement with the result of NA49
\cite{Alt:2003vb}. There is one known three star resonance in the energy range
(\ref{xirange}), namely $\Xi(1950)$ of unknown spin and parity
\cite{PDG}. However it has isospin  $I=1/2$  and therefore it cannot 
belong to $\overline{10}$.  Indeed in Ref.\cite{guz2} $\Xi(1950)$ has been attributed to the same octet as $N(1710)$. 

\begin{figure}[h]
\begin{centering}
\includegraphics[scale=0.65]{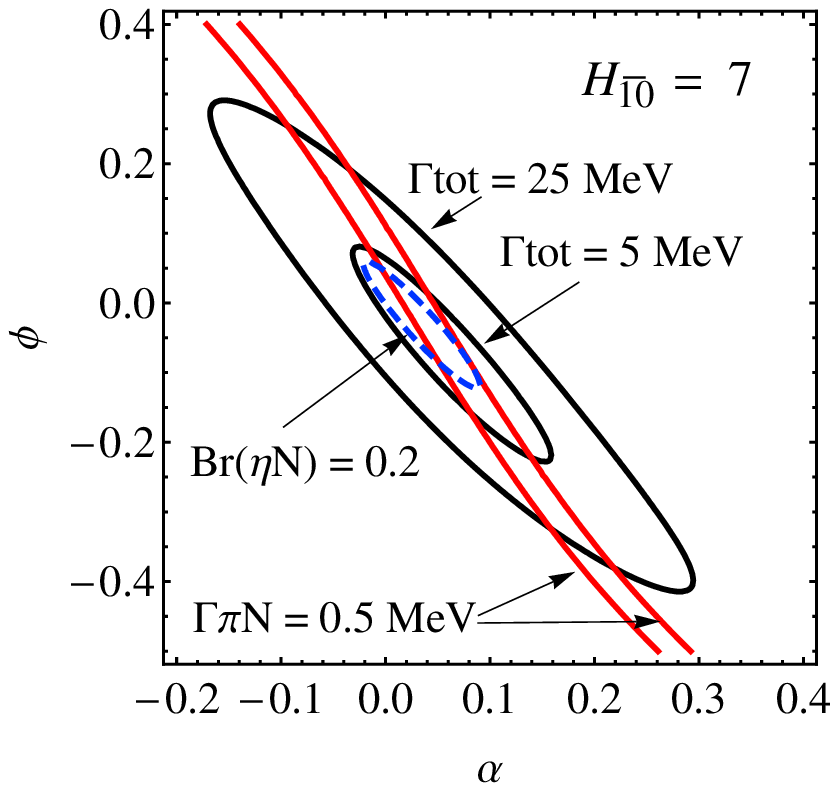}~~
\includegraphics[scale=0.65]{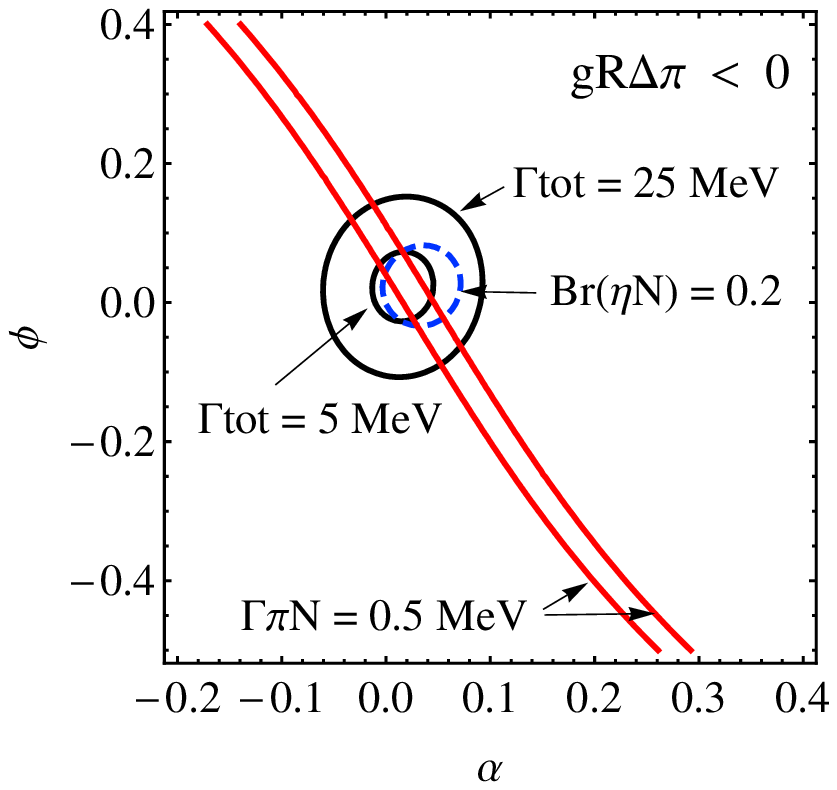}\\
\includegraphics[scale=0.65]{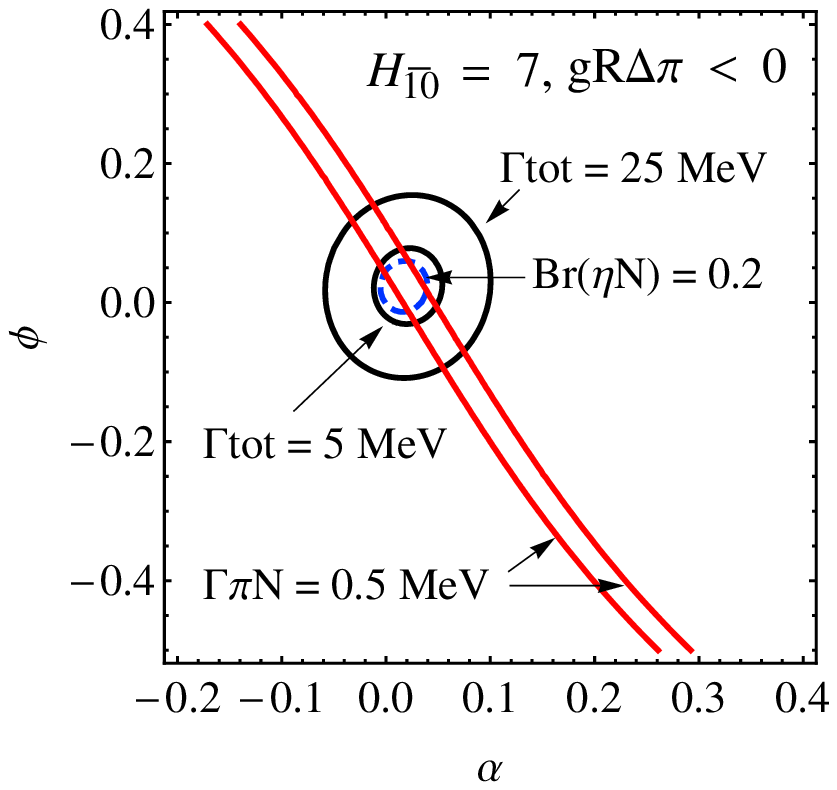}~~
\includegraphics[scale=0.65]{fig_sigma_11b.eps}\\
\par\end{centering}
\caption{Contour plots as in Fig.\ref{fig:contour} for three unconstraint
phases taking values different than in the main analysis, and for the standard
phases but for large decay width of $\Theta^{+}$.}%
\label{fig:wrong}%
\end{figure}

Our results are based on, to some extent arbitrary, assignment of the relative
signs of two reduced matrix elements defining decay constants: $H_{\overline
{10}}$ and $g_{R\Delta\pi}$. We have fixed $H_{\overline{10}}$ to be negative
($H_{\overline{10}}\sim-7$) in agreement with $\chi$QSM and $g_{R\Delta\pi}$
to be positive ($g_{R\Delta\pi}\sim30$). In Fig.\ref{fig:wrong} we show
contour plots corresponding to (\ref{conditions}) with these phases inverted.
We see that there is no common stable intersection region in the space of
mixing angles for $H_{\overline{10}}>0$ or/and $g_{R\Delta\pi}<0$. In fact
there is small allowed spot for $H_{\overline{10}}=-7$ and $g_{R\Delta\pi
}=-30$, but it is unstable for small variations of these values or conditions
(\ref{conditions}). Finally, in the same figure we present contour plot for
large width of $\Theta^{+}$ where also no allowed region of mixing angles exists.

To conclude: we have proposed a scenario in which Roper octet can mix with
putative antidecuplet of exotic baryons and predicted the properties of its
strange members $\Sigma_{\overline{10}}$ and $\Xi_{\overline{10}}$. We hope
that these estimates will be helpful in eventual experimental searches.

\renewcommand{\baselinestretch}{0.5} \renewcommand{\baselinestretch}{1.0}

\begin{acknowledgments}
The authors are grateful to M.~Amarian, D.~Diakonov, V.~Kuznetsov and V.~Petrov for discussions. The paper was partially
supported by the Polish-German cooperation agreement between Polish Academy of
Sciences (PAN) and Deutsche Forschungsgemeinschaft (DFG), and SFB TR16 (Bochum-Bonn-Giessen) of DFG.
\end{acknowledgments}

\end{document}